\documentclass[prl,aps,superscriptaddress,twocolumn,showpacs,showkeys]{revtex4-1}
\usepackage[utf8x]{inputenc}
\usepackage{microtype}
\usepackage{lmodern}
\usepackage{graphicx}
\usepackage{amsmath}
\usepackage{amsfonts}
\usepackage[colorlinks]{hyperref}

\allowdisplaybreaks[1]

\newcommand{\FM}{\lvert\mathrm{FM}\rangle}
\newcommand{\MF}{\langle\textrm{FM}\rvert}

\DeclareMathOperator{\Tr}{Tr}

\begin{document}

\title{Quantum versus classical polarons in a ferromagnetic CuO$_3$-like chain}

\author{Krzysztof Bieniasz}
\affiliation{Marian Smoluchowski Institute of Physics,
  Jagiellonian University, Reymonta 4, 30-059, Krak\'ow, Poland}
\email{krzysztof.bieniasz@uj.edu.pl}

\author{Andrzej M. Ole\'s}
\affiliation{Marian Smoluchowski Institute of Physics,
  Jagiellonian University, Reymonta 4, 30-059, Krak\'ow, Poland}
\affiliation{Max-Planck-Institut f\"ur Festk\"orperforschung,
  Heisenbergstrasse 1, D-70569 Stuttgart, Germany}

\date{\today}

\begin{abstract}
We present an exact solution for an itinerant hole added into the oxygen
orbitals of a CuO$_{3}$-like ferromagnetic chain. Using the Green's
function method, the quantum polarons obtained for the Heisenberg SU(2)
interaction between localized Cu spins are compared with the polarons
in the Ising chain. We find that magnons with large energy are
favorable towards quasiparticle existence, even in the case of
relatively modest electron-magnon coupling. We observe two quasiparticle
states with dispersion $\sim 2t$ each, which emerge from the incoherent
continuum when the exchange coupling $J$ increases. Quantum fluctuations
in the spin system modify the incoherent part of the spectrum and change
the spectral function qualitatively, beyond the bands derived from the
perturbation theory.
\end{abstract}

\maketitle

\section{Introduction}

Doping charge carriers in Mott insulators frequently leads to drastic
changes of the magnetic order and transport properties. For instance,
in colossal magnetoresistance manganites the ferromagnetic (FM) order
is accompanied by a metal-insulator transition, and appears both for
hole \cite{Tok06} and electron \cite{Ole11} doping. In contrast, the
antiferromagnetic (AF) interactions in CuO$_2$ planes is only weakened
by doping while the competition between the magnetic energy and the
hole dynamics leads to new phases of high temperature superconductors,
such as stripe \cite{Olehi} or charge order \cite{Let14}. A complete
treatment of this problem is difficult and requires a study of the
three-band model \cite{Var87,Ole87}. Therefore, simplifications
by mapping to the one-band $t$-$J$ model have been performed by several
authors \cite{Zha88,Zaa88,Fei93}.
The effective one-band model contains then next-nearest neighbor hopping
$t'$ \cite{Feine,Toh04}, and possibly even more distant hopping terms, which
influences the value of the transition temperature $T_c$ \cite{Fei96,Pav01}.

In this paper we consider a CuO$_3$-like chain (depicted in Fig.~1)
with FM exchange between localized Cu $S=1/2$ spins and
a single hole injected into the oxygen $2p$ orbitals. The chain structure
is similar to that of a CuO$_3$ chain in YBa$_2$Cu$_3$O$_7$, where the
superexchange is AF. It has been found that oxygen holes are then
delocalized and strongly correlated \cite{Ole91}. Recently, excited
states were investigated in AF CuO$_3$ chains in Sr$_2$CuO$_3$
\cite{Sch11} and an interesting interplay due to spin-orbital
entanglement \cite{Ole12} was pointed out \cite{Woh11,*Woh13}.

In case of FM ground state the single band model is also fundamentally
different from multiband models, where charge defects are generated not
in $3d$ orbitals but in $2p$ oxygen orbitals \cite{Mir12a,*Mir12b}.
This situation resembles FM semiconductors such as EuO or EuS, where
an electron with its spin aligned with the FM background moves freely,
while the one with opposite spin scatters on magnon excitations which
leads to rather complex many-body problem causing drastic modifications
of the electronic structure \cite{Nol80,*Nol81}.
In this situation as well as in the considered chain of Fig. 1, the
ground state of this model is exactly known, and the spectral properties
may be derived exactly \cite{Bie13}. Here we analyze them in detail and
we show that they include both polaron-like and scattering states when
the moving carrier interacts with magnons.

\begin{figure}[b!]
\centering
  \includegraphics[width=0.9\columnwidth]{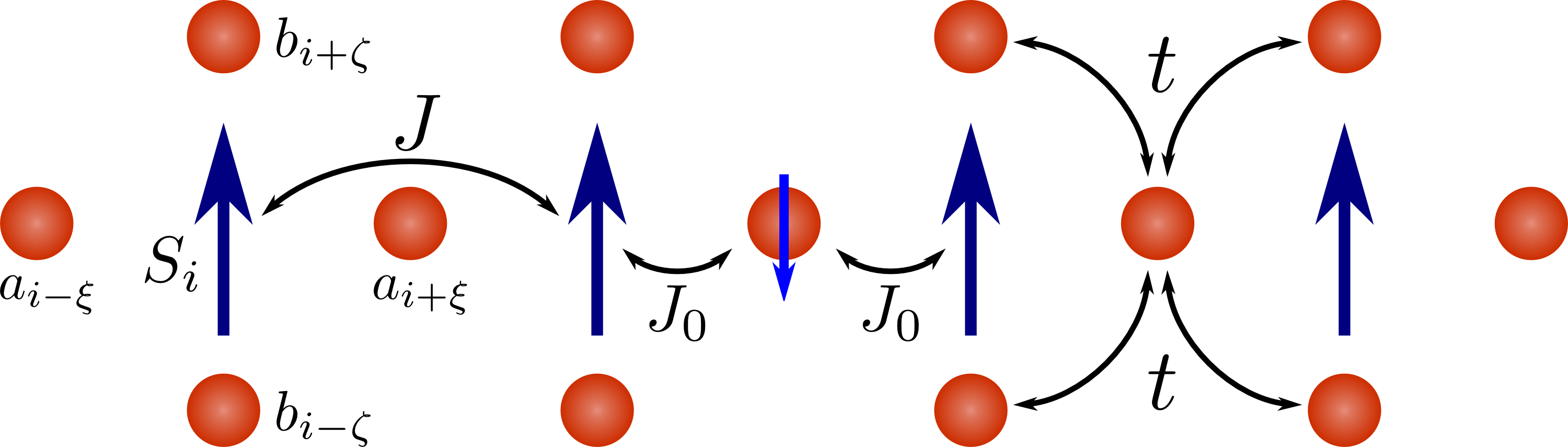}
\caption{
Schematic representation of the CuO$_3$-like chain with:
(i) exchange $J$ between neighboring Cu sites (arrows) along the chain,
(ii) Kondo coupling $J_0$ between Cu spin and the doped O spin (small arrow),
and (iii) hopping $t$ over the oxygen sites (filled circles).}
\label{fig:model}
\end{figure}

\section{The model}
\label{sec:model}

We consider a CuO$_{3}$-like FM chain, with a single $\downarrow$-spin
hole doped into either of the O($2p$) states, denoted $a$ for in-chain
orbitals and $b$ for apical orbitals (see Fig.~\ref{fig:model}).
The Cu($3d_{x^{2}-y^{2}}$) states host one
localized spin $S=1/2$ each. For simplicity, all the oxygen orbitals
are modeled as having $s$ symmetry, since the difference compared to
the $p$-$d$ model is trivial here. Further, we reduce the number of
$k$-states in the direction normal to the chain to just one, by taking
only the binding combination of apical $b$ states, i.e., for $s$
symmetry $b_{i}=(b_{i+\zeta}+b_{i-\zeta})/\sqrt{2}$. This is done to
ensure a strictly one-dimensional (1D) system and can be justified by
the fact that the antibonding states do not couple to each other and
thus do not appear in the kinetic term of the Hamiltonian.

We describe the CuO$_{3}$ chain with a $t$-$J$-like model,
\begin{equation}
\label{model}
\mathcal{H}=\mathcal{T}+\mathcal{H}_{\mathrm{S}}+\mathcal{H}_{\mathrm{K}},
\end{equation}
where the kinetic energy $\mathcal{T}$ describes the electron hopping in
the $p$-subspace, FM Heisenberg interaction $\mathcal{H}_{\mathrm{S}}$
couples the neighboring $d$ states (the constant $JS^2$ cancels the
extensive ground state energy), and a Kondo-like $p$-$d$ exchange term
$\mathcal{H}_{\mathrm{K}}$ which couples the two subspaces:
\begin{subequations}
  \label{eq:hamilt1}
  \begin{align}
    \label{eq:t}
    \mathcal{T} &= -t\sum_{i\sigma}
    \left[(a_{i+\xi,\sigma}^{\dag} + a_{i-\xi,\sigma}^{\dag})
      b_{i\sigma} + \text{H.c.}\right],\\
    \label{eq:hs}
    \mathcal{H}_{\mathrm{S}} &= -J\sum_{i}
    (\mathbf{S}_{i}\cdot\mathbf{S}_{i+1}-S^{2}),\\
    \label{eq:hk}
    \mathcal{H}_{\mathrm{K}} &= J_{0}\sum_{i}
    \left(\mathbf{s}_{i+\xi}^{a}+\mathbf{s}_{i-\xi}^{a}
      +\mathbf{s}_{i}^{b}\right) \cdot \mathbf{S}_{i},
  \end{align}
\end{subequations}
where all the energy parameters are taken as positive, i.e., $t>0$,
$J>0$ and $J_0>0$. Oxygen site spin operators $\mathbf{s}_{j}^{m}$
are labeled by the site index $j$ and the orbital index $m$
serves as a reminder which of the $p$ orbitals the site corresponds to.
These operators are later
expressed in standard fermionic representation for $s=1/2$ spins.
We also consider two symmetries in the $\mathcal{H}_{\mathrm{S}}$ term:
$(i)$ the SU(2) symmetry corresponding to the Heisenberg model
\eqref{eq:hs}, and
$(ii)$ the $Z_2$ symmetry realized in the Ising model, where the scalar
product in \eqref{eq:hs} is replaced by the Ising term,
$\mathbf{S}_{i}\cdot\mathbf{S}_{i+1}\rightarrow S_{i}^zS_{i+1}^z$.

To proceed, one performs a Fourier transformation (FT) and introduces
a convenient matrix notation, which leads to the following
representation of the Hamiltonian \eqref{eq:hamilt1} in the
$p$-orbital basis:
\begin{subequations}
  \label{eq:matrix}
  \begin{align}
    \label{tmatrix}
    \mathbb{T}(k) &=
    \begin{pmatrix}
      0 & \epsilon_{k}^{}\\
      \epsilon_{k}^{*} & 0
    \end{pmatrix},\\
    \label{eq:vmatrix}
    \mathbb{V}(q) &=
    \begin{pmatrix}
      \cos(q/2) & 0\\
      0 & 1/2
    \end{pmatrix},
  \end{align}
\end{subequations}
where $\mathbb{V}(q)$ represents $\mathcal{H}_{\mathrm{K}}$ and
$\epsilon_{k}^{}=-2t\cos(k/2)$ is the dispersion relation for a
bare itinerant hole. $\mathcal{H}_{\mathrm{S}}$ is treated
separately by noting that in the $p$-orbital basis it has an
identity representation, and two eigenvalues
corresponding to the eigenstates spanning the magnetic subspace:
\begin{align}
  \label{eq:Hh0}
  \mathcal{H}_{\mathrm{S}} \FM &= 0 \FM,\\
  \label{eq:HhSq}
  \mathcal{H}_{\mathrm{S}} S_{q}^{-} \FM &= \Omega_{q} S_{q}^{-} \FM,
\end{align}
where
\begin{equation}
  \label{eq:omq}
  \Omega_{q}=
  \begin{cases}
    4JS\sin^{2}(q/2), & \text{for Heisenberg $\mathcal{H}_{\mathrm{S}}$},\\
    2JS, & \text{for Ising $\mathcal{H}_{\mathrm{S}}$},
  \end{cases}
\end{equation}
is the magnon dispersion relation and the FT spin operator is
defined as $S_{q}^-=\frac{1}{N}\sum_{i} e^{-iqR_{i}}S_{i}^{-}$.

The problem outlined above can be solved exactly by Green's
functions \cite{berciu09,Mir12a}, defined as
the matrix representation of the resolvent operator
$\mathcal{G}(\omega) = [\omega-\mathcal{H}+i\eta]^{-1}$ in the state
with one hole,
\begin{equation}
  \label{eq:green}
  \mathbb{G}_{\mu\nu}(k,\omega) =
\MF\mu_{k\downarrow}\mathcal{G}(\omega)\nu_{k\downarrow}^{\dag}\FM,
\end{equation}
where $\mu,\nu\in\{a,b\}$ are indices running over the set of all orbitals taking
part in the hole dynamics.

We separate the Hamiltonian \eqref{eq:hamilt1} into the bare part
$\mathcal{H}_{0}=\mathcal{T}+\mathcal{H}_{\mathrm{S}}$ and the
interaction $\mathcal{V}=\mathcal{H}_{\mathrm{K}}$. We define the
free Green's function $\mathbb{G}_{0}(k,\omega)$ as the Green's
function corresponding to $\mathcal{H}_{0}$, and the full
Green's function corresponds to the complete Hamiltonian
\eqref{eq:hamilt1}. Next, we perform a Dyson expansion of the full Green's function. Due
to the very constrained magnetic Hilbert space consisting of just two
distinct states, after performing the expansion twice the equations
close, and one can express $\mathbb{G}(k,\omega)$ solely in terms of
the free Green's function $\mathbb{G}_{0}(k,\omega)$.

Due to very limited space available for this article, we do not present
here any details of the rather tedious derivation. They may be found
in the original paper \cite{Bie13}, where the full derivation of the
spectral function is presented. Here we give only the final result
needed for the numerical analysis presented below. The full
Green's function can thus be expressed in the following way:
\begin{eqnarray}
    \label{eq:G2}
    \mathbb{G}(k,\omega) &=&
    \Big[\left[\mathbb{G}_{0}(k,\omega)\mathbb{Q}_{+}(k,\omega)\right]^{-1}
    \nonumber \\
    &&-2J_{0}S \left[\mathbb{I}-\mathbb{M}^{-1}(k,\omega)\right] \Big]^{-1},\\
    \label{eq:M}
    \mathbb{M}(k,\omega) &=& \mathbb{I} +\mathbb{G}_{cc}(k,\omega)
    -\mathbb{G}_{cs}(k,\omega)\nonumber \\
    &&\times[\mathbb{I}+\mathbb{G}_{ss}(k,\omega)]^{-1}
    \mathbb{G}_{sc}(k,\omega)\,,\\
    \label{eq:gcc}
  \mathbb{G}_{\alpha\beta} &=& \frac{J_{0}}{N}\sum_{q}
  \mathbb{U}_{\alpha}(q) \mathbb{G}_{0}(k-q,\omega-\Omega_{q})\nonumber \\
    &&\times\mathbb{Q}_{-}(k-q,\omega-\Omega_{q}) \mathbb{U}_{\beta}(q)\,,
\end{eqnarray}
where
\begin{align}
  \label{eq:uab}
  \mathbb{U}_{\mu}(q) &=
  \begin{cases}
    \mathbb{V}(q), & \mu=c,\\
    \bar{\mathbb{V}}(q)=
    \left(
    \begin{smallmatrix}
      \sin(q/2) & 0\\
      0 & 1/2
    \end{smallmatrix}
    \right), & \mu=s,
  \end{cases}\\
  \label{eq:qmatrix}
  \mathbb{Q}_{\pm}(k,\omega) &= \big[\mathbb{I}
  \pm J_{0}S\mathbb{V}(0) \mathbb{G}_{0}(k,\omega)\big]^{-1}.
\end{align}

Having calculated the Green's function \eqref{eq:G2}, we extract
from it the physical information in the form of the traced spectral
function,
\begin{equation}
  \label{eq:spectral}
  A(k,\omega) = -\frac{1}{2\pi} \Im \left[
    \Tr \mathbb{G}(k,\omega)
  \right],
\end{equation}
which is useful in the interpretation of the spectra derived
from photoelectron spectroscopy experiments. In practice, throughout this article
we plot $\tanh[A(k,\omega)]$ to bring out the low amplitude part of the spectra.

\section{Results and discussion}
\label{sec:results}

In our previous paper \cite{Bie13} we reported on the evolution of the
spectral function \eqref{eq:spectral} with increasing electron-magnon
coupling strength, characterized by the parameter $J_{0}$. We have
found that with increasing value of $J_{0}$, the spectral function
changes from having two states, corresponding to the two branches of
the free hole dispersion, to the five polaronic states, whose nature we
explored in detail using perturbation theory \cite{Bie13}.
We also benchmarked those results against the mean field (MF)
approximation, finding that MF works well for weak coupling, while for
strong coupling it highly underrates the QP binding energy, even though
the quantitative results (i.e., the band shapes) are predicted
surprisingly well.

\begin{figure}[t!]
  \centering
  \includegraphics[width=\columnwidth]{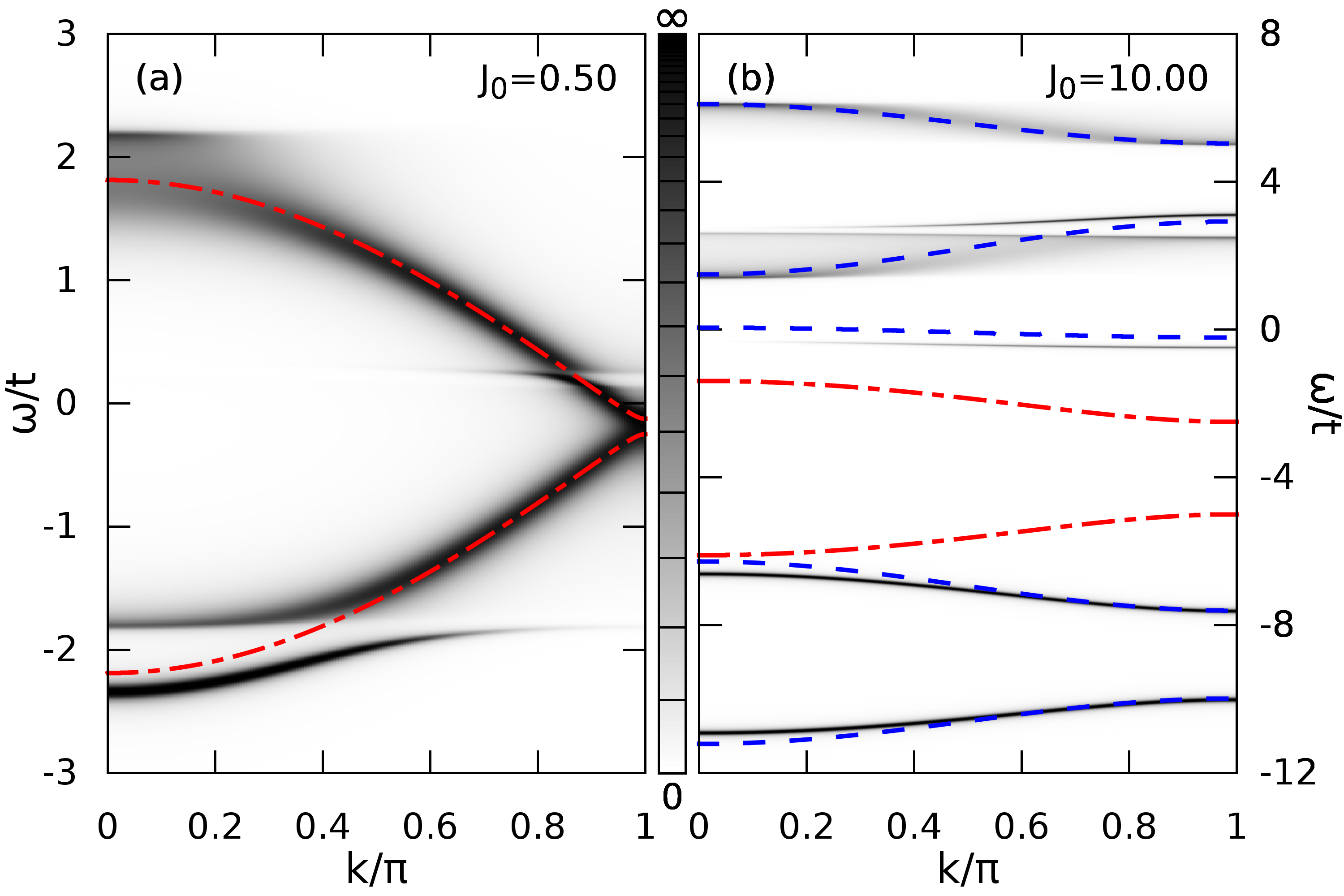}
\caption{Spectral functions compared with the MF solution
(red dash-dotted lines) and with the perturbation expansion
(blue dashed lines), as obtained for:
(a) $J_0=0.5t$, and
(b) $J_0=10t$.
Note the highly nonlinear tanh-scale, with tics spaced every 0.1.
Parameters: $J=0.05t$, $\eta=0.02t$. }
  \label{fig:summary}
\end{figure}

Figure~\ref{fig:summary} presents a very condensed summary of our
previous results \cite{Bie13}. Already for $J_{0}=0.5t$ one can see
that new bands and gaps develop in the spectra. However, MF
approximation (red dash-dotted lines) still works quite well
predicting the localization of the highest density in the graph.
On the other hand, for an extremely high value of $J_{0}=10t$, MF
breaks down completely, while the perturbation expansion in $t$ and
$J$ (blue dashed lines) replicates the maxima of the spectral function
quite well, allowing one to identify those states as polaron-like.

\begin{figure}[t!]
\centering
  \includegraphics[width=\columnwidth]{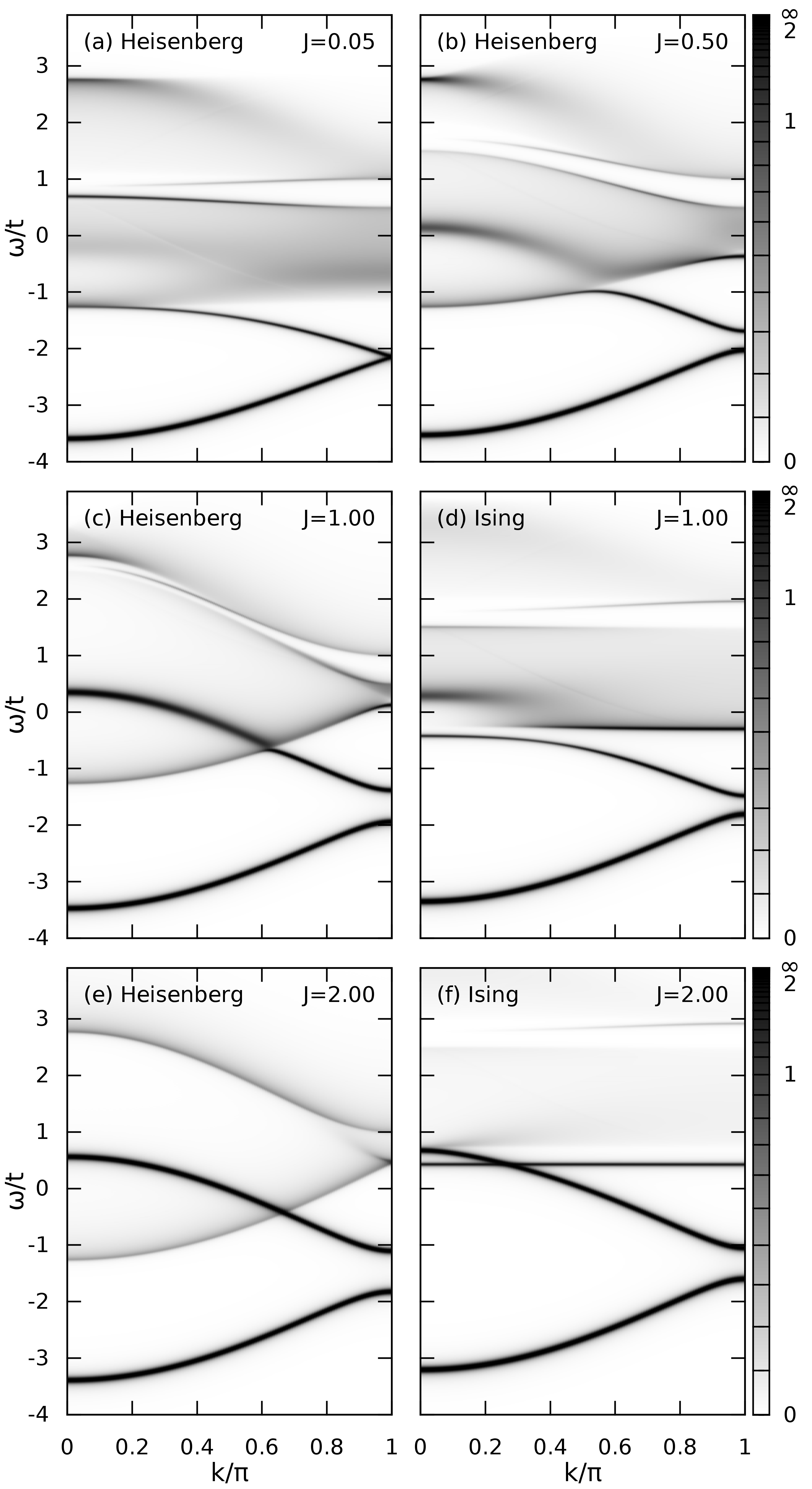}
\caption{
Spectral function density maps for a broad range of $J$ values
obtained for:
Heisenberg (a-c,e) and
Ising (d,f) spin interaction.
Note the highly nonlinear tanh-scale (right), with tics spaced every 0.1,
used to display the features with low intensity.
Parameters: $J_{0}=2t$ and $\eta=0.02t$.}
  \label{fig:layout}
\end{figure}

In the present work we focus on the importance of the magnon energy,
i.e., on establishing how the value of the parameter $J$ influences the
spectra. Figure~\ref{fig:layout} presents the evolution of the spectral
function when increasing $J$ in the range $0.05t$ to $2.0t$.
The value of $J_0$ is set at $2.0t$, chosen so that at small $J$ the
spectrum already shows some of the polaronic features, however the bands
are not yet fully developed. In fact, Fig.~\ref{fig:layout}a shows this
for $J=0.05$, the value used in our previous research, which serves here
as a reference state. In this graph a well developed lowest branch can
be recognized, and another one just above it, still emerging from the
incoherent continuum; the incoherent part is divided into two areas,
separated by a small gap and well defined boundaries. Increasing $J$ to
$0.5t$ (Fig.~\ref{fig:layout}b) one notices that the second band starts
to mix with the incoherent spectra and a gap opens between the two
branches. At $J=t$ (Fig.~\ref{fig:layout}c) a new structure fades in
from the incoherence, while most of the background disappears. Finally,
at $J=2t$ (Fig.~\ref{fig:layout}e) the second branch is fully developed,
while the incoherent part has mostly collapsed into a pair of ``ghost''
bands.

The spectral functions presented above have a number of noteworthy
features. Firstly, compared to the strong electron-magnon coupling,
even quite modest magnon energies (i.e., both $J_{0}$ and $J$ are small
and neither is the leading term) suppress the incoherent part of the
spectra and aid the development of QP bands. For example, instead of
three incoherent and rather complex features for $J_0=10t$, $J=0.05t$
(Fig.~\ref{fig:summary}b) one gets just two well defined and low
amplitude bands for $J_0=2t$, $J=2t$ (Fig.~\ref{fig:layout}e), whose
spectral weight decreases with increasing $J$. Those two branches
correspond to the two incoherent regions mentioned before, which seems
to suggest that coherence is directly linked to magnon energy.

Finally, let us consider the case of Ising spin exchange and compare it
with Heisenberg $\mathcal{H}_{\mathrm{S}}$.
Figures~\ref{fig:layout}d~\&~\ref{fig:layout}f present the spectral
functions for the Ising case, for $J=1t$ and $J=2t$ respectively. For
very small $J=0.05t$ (not shown), local $J_0$ term dominates and there
is practically no difference between Heisenberg and Ising magnons.
On the other hand, for stronger exchange interaction $J$ the picture
changes quite substantially. One notices that two QP bands exist in
the Ising case for $J=2t$ (Fig. \ref{fig:summary}f), with a slightly
higher (although $J$-independent) binding energy and a very similar
dispersion as in the Heisenberg case (Fig. \ref{fig:summary}e).
However, the incoherent part of the spectrum changes far less
dramatically for Ising than it does for Heisenberg interaction
$\mathcal{H}_{\mathrm{S}}$ when increasing $J$. Its spectral weight diminishes and the
different branches move to higher energies as dictated by the value of
$J$, but they never collapse into two states with energies increased by
$J_0$, as seen for the Heisenberg case. The reason for this can be
revealed using the perturbation theory.

\begin{figure}[t!]
\centering
  \includegraphics[width=\columnwidth]{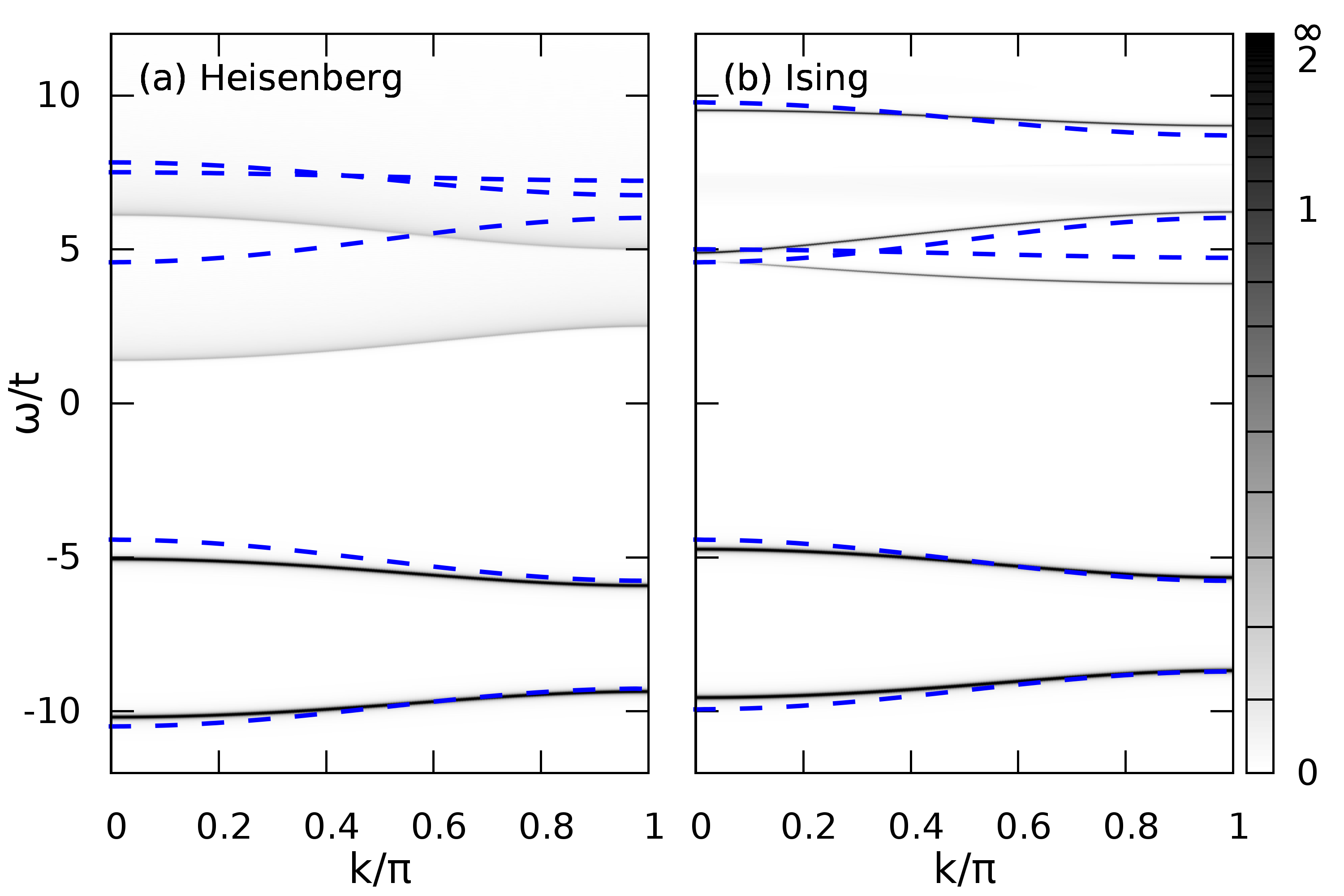}
\caption{
Spectral weight distribution as obtained in perturbation theory (blue
dashed lines) compared with the exact solution (shaded) for:
(a) Heisenberg, and
(b) Ising exchange between localized spins.
Note the nonlinear tanh-scale.
Parameters: $J_{0}=10t$ and $J=5t$.}
  \label{fig:perturb}
\end{figure}

Figure~\ref{fig:perturb} shows the results of perturbation expansion,
as outlined in \cite{Bie13}, against the exact results for very big
$J_{0}=10t$ and big $J=5t$, for both the Heisenberg and Ising cases.
This reveals some of the peculiarities in the interplay between those
two parameters. First, note that for the Ising case the incoherent
(upper) part of the spectrum is noticeably higher than for the
Heisenberg case, while no such shift can be observed for the QP bands.
This suggests that the quantum spin fluctuations affect only the
incoherent spectra, while they are irrelevant for the QPs. Second,
only two branches are seen for the Heisenberg spectrum, while there
are three distinct features for Ising case. Thus, it is clear that
while the perturbation expansion works well for Ising and breaks down
for the Heisenberg case. This breakdown depends strongly on the
value of $J$, with the perturbation solution going
gradually out of tune with the exact one with increasing $J$.
This demonstrates a drastic redistribution of the spectral weight
due to the mixing of incoherent processes.

Summarizing, we have shown that the perturbation expansion reproduces
the spectra obtained in the Ising limit, while quantum fluctuations
modify the incoherent part of the spectrum and generate two dispersive
states. This demonstrates that quantum spin fluctuations strongly
affect the incoherent part of the spectra, while they almost do not
contribute to the quasiparticle part at low energy.

\acknowledgments
We thank Mona Berciu for stimulating discussions.
We kindly acknowledge financial support by the Polish National
Science Center (NCN) under Project No.~2012/04/A/ST3/00331.

\bibliographystyle{apsrev4-1}
\bibliography{bib}

\begin{thebibliography}{24}%
\makeatletter
\providecommand \@ifxundefined [1]{%
 \@ifx{#1\undefined}
}%
\providecommand \@ifnum [1]{%
 \ifnum #1\expandafter \@firstoftwo
 \else \expandafter \@secondoftwo
 \fi
}%
\providecommand \@ifx [1]{%
 \ifx #1\expandafter \@firstoftwo
 \else \expandafter \@secondoftwo
 \fi
}%
\providecommand \natexlab [1]{#1}%
\providecommand \enquote  [1]{``#1''}%
\providecommand \bibnamefont  [1]{#1}%
\providecommand \bibfnamefont [1]{#1}%
\providecommand \citenamefont [1]{#1}%
\providecommand \href@noop [0]{\@secondoftwo}%
\providecommand \href [0]{\begingroup \@sanitize@url \@href}%
\providecommand \@href[1]{\@@startlink{#1}\@@href}%
\providecommand \@@href[1]{\endgroup#1\@@endlink}%
\providecommand \@sanitize@url [0]{\catcode `\\12\catcode `\$12\catcode
  `\&12\catcode `\#12\catcode `\^12\catcode `\_12\catcode `\%12\relax}%
\providecommand \@@startlink[1]{}%
\providecommand \@@endlink[0]{}%
\providecommand \url  [0]{\begingroup\@sanitize@url \@url }%
\providecommand \@url [1]{\endgroup\@href {#1}{\urlprefix }}%
\providecommand \urlprefix  [0]{URL }%
\providecommand \Eprint [0]{\href }%
\providecommand \doibase [0]{http://dx.doi.org/}%
\providecommand \selectlanguage [0]{\@gobble}%
\providecommand \bibinfo  [0]{\@secondoftwo}%
\providecommand \bibfield  [0]{\@secondoftwo}%
\providecommand \translation [1]{[#1]}%
\providecommand \BibitemOpen [0]{}%
\providecommand \bibitemStop [0]{}%
\providecommand \bibitemNoStop [0]{.\EOS\space}%
\providecommand \EOS [0]{\spacefactor3000\relax}%
\providecommand \BibitemShut  [1]{\csname bibitem#1\endcsname}%
\let\auto@bib@innerbib\@empty
\bibitem [{\citenamefont {Tokura}(2006)}]{Tok06}%
  \BibitemOpen
  \bibfield  {author} {\bibinfo {author} {\bibfnamefont {Y.}~\bibnamefont
  {Tokura}},\ }\href {\doibase 10.1088/0034-4885/69/3/R06} {\bibfield
  {journal} {\bibinfo  {journal} {Rep. Prog. Phys.}\ }\textbf {\bibinfo
  {volume} {69}},\ \bibinfo {pages} {797} (\bibinfo {year} {2006})}\BibitemShut
  {NoStop}%
\bibitem [{\citenamefont {Ole\'s}\ and\ \citenamefont
  {Khaliullin}(2011)}]{Ole11}%
  \BibitemOpen
  \bibfield  {author} {\bibinfo {author} {\bibfnamefont {A.~M.}\ \bibnamefont
  {Ole\'s}}\ and\ \bibinfo {author} {\bibfnamefont {G.}~\bibnamefont
  {Khaliullin}},\ }\href {\doibase 10.1103/PhysRevB.84.214414} {\bibfield
  {journal} {\bibinfo  {journal} {Phys. Rev. B}\ }\textbf {\bibinfo {volume}
  {84}},\ \bibinfo {pages} {214414} (\bibinfo {year} {2011})}\BibitemShut
  {NoStop}%
\bibitem [{\citenamefont {Ole\'s}(2012{\natexlab{a}})}]{Olehi}%
  \BibitemOpen
  \bibfield  {author} {\bibinfo {author} {\bibfnamefont {A.~M.}\ \bibnamefont
  {Ole\'s}},\ }\href {http://przyrbwn.icm.edu.pl/APP/PDF/121/a121z4p108.pdf}
  {\bibfield  {journal} {\bibinfo  {journal} {Acta Phys. Polon. A}\ }\textbf
  {\bibinfo {volume} {121}},\ \bibinfo {pages} {752} (\bibinfo {year}
  {2012}{\natexlab{a}})}\BibitemShut {NoStop}%
\bibitem [{\citenamefont {Le~Tacon}\ \emph {et~al.}(2014)\citenamefont
  {Le~Tacon} \emph {et~al.}}]{Let14}%
  \BibitemOpen
  \bibfield  {author} {\bibinfo {author} {\bibfnamefont {M.}~\bibnamefont
  {Le~Tacon}} \emph {et~al.},\ }\href {\doibase 10.1038/nphys2805} {\bibfield
  {journal} {\bibinfo  {journal} {Nat. Phys.}\ }\textbf {\bibinfo {volume}
  {10}},\ \bibinfo {pages} {52} (\bibinfo {year} {2014})}\BibitemShut {NoStop}%
\bibitem [{\citenamefont {Varma}\ \emph {et~al.}(1987)\citenamefont {Varma},
  \citenamefont {Schmitt-Rink},\ and\ \citenamefont {Abrahams}}]{Var87}%
  \BibitemOpen
  \bibfield  {author} {\bibinfo {author} {\bibfnamefont {C.~M.}\ \bibnamefont
  {Varma}}, \bibinfo {author} {\bibfnamefont {S.}~\bibnamefont {Schmitt-Rink}},
  \ and\ \bibinfo {author} {\bibfnamefont {E.}~\bibnamefont {Abrahams}},\
  }\href {\doibase 10.1016/0038-1098(87)90407-8} {\bibfield  {journal}
  {\bibinfo  {journal} {Solid State Commun.}\ }\textbf {\bibinfo {volume}
  {62}},\ \bibinfo {pages} {681} (\bibinfo {year} {1987})}\BibitemShut
  {NoStop}%
\bibitem [{\citenamefont {Ole\'s}\ \emph {et~al.}(1987)\citenamefont {Ole\'s},
  \citenamefont {Zaanen},\ and\ \citenamefont {Fulde}}]{Ole87}%
  \BibitemOpen
  \bibfield  {author} {\bibinfo {author} {\bibfnamefont {A.~M.}\ \bibnamefont
  {Ole\'s}}, \bibinfo {author} {\bibfnamefont {J.}~\bibnamefont {Zaanen}}, \
  and\ \bibinfo {author} {\bibfnamefont {P.}~\bibnamefont {Fulde}},\ }\href
  {\doibase 10.1016/0378-4363(87)90205-1} {\bibfield  {journal} {\bibinfo
  {journal} {Physica B+C}\ }\textbf {\bibinfo {volume} {148}},\ \bibinfo
  {pages} {260} (\bibinfo {year} {1987})}\BibitemShut {NoStop}%
\bibitem [{\citenamefont {Zhang}\ and\ \citenamefont {Rice}(1988)}]{Zha88}%
  \BibitemOpen
  \bibfield  {author} {\bibinfo {author} {\bibfnamefont {F.~C.}\ \bibnamefont
  {Zhang}}\ and\ \bibinfo {author} {\bibfnamefont {T.~M.}\ \bibnamefont
  {Rice}},\ }\href {\doibase 10.1103/PhysRevB.37.3759} {\bibfield  {journal}
  {\bibinfo  {journal} {Phys. Rev.~B}\ }\textbf {\bibinfo {volume} {37}},\
  \bibinfo {pages} {3759} (\bibinfo {year} {1988})}\BibitemShut {NoStop}%
\bibitem [{\citenamefont {Zaanen}\ and\ \citenamefont {Ole\'s}(1988)}]{Zaa88}%
  \BibitemOpen
  \bibfield  {author} {\bibinfo {author} {\bibfnamefont {J.}~\bibnamefont
  {Zaanen}}\ and\ \bibinfo {author} {\bibfnamefont {A.~M.}\ \bibnamefont
  {Ole\'s}},\ }\href {\doibase 10.1103/PhysRevB.37.9423} {\bibfield  {journal}
  {\bibinfo  {journal} {Phys. Rev.~B}\ }\textbf {\bibinfo {volume} {37}},\
  \bibinfo {pages} {9423} (\bibinfo {year} {1988})}\BibitemShut {NoStop}%
\bibitem [{\citenamefont {Feiner}(1993)}]{Fei93}%
  \BibitemOpen
  \bibfield  {author} {\bibinfo {author} {\bibfnamefont {L.~F.}\ \bibnamefont
  {Feiner}},\ }\href {\doibase 10.1103/PhysRevB.48.16857} {\bibfield  {journal}
  {\bibinfo  {journal} {Phys. Rev.~B}\ }\textbf {\bibinfo {volume} {48}},\
  \bibinfo {pages} {16857} (\bibinfo {year} {1993})}\BibitemShut {NoStop}%
\bibitem [{\citenamefont {Feiner}\ \emph
  {et~al.}(1996{\natexlab{a}})\citenamefont {Feiner}, \citenamefont
  {Jefferson},\ and\ \citenamefont {Raimondi}}]{Feine}%
  \BibitemOpen
  \bibfield  {author} {\bibinfo {author} {\bibfnamefont {L.~F.}\ \bibnamefont
  {Feiner}}, \bibinfo {author} {\bibfnamefont {J.~H.}\ \bibnamefont
  {Jefferson}}, \ and\ \bibinfo {author} {\bibfnamefont {R.}~\bibnamefont
  {Raimondi}},\ }\href {\doibase 10.1103/PhysRevB.53.8751} {\bibfield
  {journal} {\bibinfo  {journal} {Phys. Rev.~B}\ }\textbf {\bibinfo {volume}
  {53}},\ \bibinfo {pages} {8751} (\bibinfo {year}
  {1996}{\natexlab{a}})}\BibitemShut {NoStop}%
\bibitem [{\citenamefont {Tohyama}(2004)}]{Toh04}%
  \BibitemOpen
  \bibfield  {author} {\bibinfo {author} {\bibfnamefont {T.}~\bibnamefont
  {Tohyama}},\ }\href {\doibase 10.1103/PhysRevB.70.174517} {\bibfield
  {journal} {\bibinfo  {journal} {Phys. Rev. B}\ }\textbf {\bibinfo {volume}
  {70}},\ \bibinfo {pages} {174517} (\bibinfo {year} {2004})}\BibitemShut
  {NoStop}%
\bibitem [{\citenamefont {Feiner}\ \emph
  {et~al.}(1996{\natexlab{b}})\citenamefont {Feiner}, \citenamefont
  {Jefferson},\ and\ \citenamefont {Raimondi}}]{Fei96}%
  \BibitemOpen
  \bibfield  {author} {\bibinfo {author} {\bibfnamefont {L.~F.}\ \bibnamefont
  {Feiner}}, \bibinfo {author} {\bibfnamefont {J.~H.}\ \bibnamefont
  {Jefferson}}, \ and\ \bibinfo {author} {\bibfnamefont {R.}~\bibnamefont
  {Raimondi}},\ }\href {\doibase 10.1103/PhysRevLett.76.4939} {\bibfield
  {journal} {\bibinfo  {journal} {Phys. Rev. Lett.}\ }\textbf {\bibinfo
  {volume} {76}},\ \bibinfo {pages} {4939} (\bibinfo {year}
  {1996}{\natexlab{b}})}\BibitemShut {NoStop}%
\bibitem [{\citenamefont {Pavarini}\ \emph {et~al.}(2007)\citenamefont
  {Pavarini}, \citenamefont {Dasgupta}, \citenamefont {Saha-Dasgupta},
  \citenamefont {Jepsen},\ and\ \citenamefont {Andersen}}]{Pav01}%
  \BibitemOpen
  \bibfield  {author} {\bibinfo {author} {\bibfnamefont {E.}~\bibnamefont
  {Pavarini}}, \bibinfo {author} {\bibfnamefont {I.}~\bibnamefont {Dasgupta}},
  \bibinfo {author} {\bibfnamefont {T.}~\bibnamefont {Saha-Dasgupta}}, \bibinfo
  {author} {\bibfnamefont {O.}~\bibnamefont {Jepsen}}, \ and\ \bibinfo {author}
  {\bibfnamefont {O.~K.}\ \bibnamefont {Andersen}},\ }\href {\doibase
  10.1103/PhysRevLett.87.047003} {\bibfield  {journal} {\bibinfo  {journal}
  {Phys. Rev. Lett.}\ }\textbf {\bibinfo {volume} {87}},\ \bibinfo {pages}
  {047003} (\bibinfo {year} {2007})}\BibitemShut {NoStop}%
\bibitem [{\citenamefont {Ole\'s}\ and\ \citenamefont {Grzelka}(1991)}]{Ole91}%
  \BibitemOpen
  \bibfield  {author} {\bibinfo {author} {\bibfnamefont {A.~M.}\ \bibnamefont
  {Ole\'s}}\ and\ \bibinfo {author} {\bibfnamefont {W.}~\bibnamefont
  {Grzelka}},\ }\href {\doibase 10.1103/PhysRevB.44.9531} {\bibfield  {journal}
  {\bibinfo  {journal} {Phys. Rev.~B}\ }\textbf {\bibinfo {volume} {44}},\
  \bibinfo {pages} {9531} (\bibinfo {year} {1991})}\BibitemShut {NoStop}%
\bibitem [{\citenamefont {Schlappa}\ \emph {et~al.}(2011)\citenamefont
  {Schlappa} \emph {et~al.}}]{Sch11}%
  \BibitemOpen
  \bibfield  {author} {\bibinfo {author} {\bibfnamefont {J.}~\bibnamefont
  {Schlappa}} \emph {et~al.},\ }\href {\doibase 10.1038/nature10974} {\bibfield
   {journal} {\bibinfo  {journal} {Nature}\ }\textbf {\bibinfo {volume}
  {485}},\ \bibinfo {pages} {82} (\bibinfo {year} {2011})}\BibitemShut
  {NoStop}%
\bibitem [{\citenamefont {Ole\'s}(2012{\natexlab{b}})}]{Ole12}%
  \BibitemOpen
  \bibfield  {author} {\bibinfo {author} {\bibfnamefont {A.~M.}\ \bibnamefont
  {Ole\'s}},\ }\href {\doibase 10.1088/0953-8984/24/31/313201} {\bibfield
  {journal} {\bibinfo  {journal} {J. Phys.: Condens. Matter}\ }\textbf
  {\bibinfo {volume} {24}},\ \bibinfo {pages} {313201} (\bibinfo {year}
  {2012}{\natexlab{b}})}\BibitemShut {NoStop}%
\bibitem [{\citenamefont {Wohlfeld}\ \emph {et~al.}(2011)\citenamefont
  {Wohlfeld}, \citenamefont {Daghofer}, \citenamefont {Nishimoto},
  \citenamefont {Khaliullin},\ and\ \citenamefont {van~den Brink}}]{Woh11}%
  \BibitemOpen
  \bibfield  {author} {\bibinfo {author} {\bibfnamefont {K.}~\bibnamefont
  {Wohlfeld}}, \bibinfo {author} {\bibfnamefont {M.}~\bibnamefont {Daghofer}},
  \bibinfo {author} {\bibfnamefont {S.}~\bibnamefont {Nishimoto}}, \bibinfo
  {author} {\bibfnamefont {G.}~\bibnamefont {Khaliullin}}, \ and\ \bibinfo
  {author} {\bibfnamefont {J.}~\bibnamefont {van~den Brink}},\ }\href {\doibase
  10.1103/PhysRevLett.107.147201} {\bibfield  {journal} {\bibinfo  {journal}
  {Phys. Rev. Lett.}\ }\textbf {\bibinfo {volume} {107}},\ \bibinfo {pages}
  {147201} (\bibinfo {year} {2011})}\BibitemShut {NoStop}%
\bibitem [{\citenamefont {Wohlfeld}\ \emph {et~al.}(2013)\citenamefont
  {Wohlfeld}, \citenamefont {Nishimoto}, \citenamefont {Haverkort},\ and\
  \citenamefont {van~den Brink}}]{Woh13}%
  \BibitemOpen
  \bibfield  {author} {\bibinfo {author} {\bibfnamefont {K.}~\bibnamefont
  {Wohlfeld}}, \bibinfo {author} {\bibfnamefont {S.}~\bibnamefont {Nishimoto}},
  \bibinfo {author} {\bibfnamefont {M.~W.}\ \bibnamefont {Haverkort}}, \ and\
  \bibinfo {author} {\bibfnamefont {J.}~\bibnamefont {van~den Brink}},\ }\href
  {\doibase 10.1103/PhysRevB.88.195138} {\bibfield  {journal} {\bibinfo
  {journal} {Phys. Rev. B}\ }\textbf {\bibinfo {volume} {88}},\ \bibinfo
  {pages} {195138} (\bibinfo {year} {2013})}\BibitemShut {NoStop}%
\bibitem [{\citenamefont {M\"oller}\ \emph
  {et~al.}(2012{\natexlab{a}})\citenamefont {M\"oller}, \citenamefont
  {Sawatzky},\ and\ \citenamefont {Berciu}}]{Mir12a}%
  \BibitemOpen
  \bibfield  {author} {\bibinfo {author} {\bibfnamefont {M.}~\bibnamefont
  {M\"oller}}, \bibinfo {author} {\bibfnamefont {G.~A.}\ \bibnamefont
  {Sawatzky}}, \ and\ \bibinfo {author} {\bibfnamefont {M.}~\bibnamefont
  {Berciu}},\ }\href {\doibase 10.1103/PhysRevLett.108.216403} {\bibfield
  {journal} {\bibinfo  {journal} {Phys. Rev. Lett.}\ }\textbf {\bibinfo
  {volume} {108}},\ \bibinfo {pages} {216403} (\bibinfo {year}
  {2012}{\natexlab{a}})}\BibitemShut {NoStop}%
\bibitem [{\citenamefont {M\"oller}\ \emph
  {et~al.}(2012{\natexlab{b}})\citenamefont {M\"oller}, \citenamefont
  {Sawatzky},\ and\ \citenamefont {Berciu}}]{Mir12b}%
  \BibitemOpen
  \bibfield  {author} {\bibinfo {author} {\bibfnamefont {M.}~\bibnamefont
  {M\"oller}}, \bibinfo {author} {\bibfnamefont {G.~A.}\ \bibnamefont
  {Sawatzky}}, \ and\ \bibinfo {author} {\bibfnamefont {M.}~\bibnamefont
  {Berciu}},\ }\href {\doibase 10.1103/PhysRevB.86.075128} {\bibfield
  {journal} {\bibinfo  {journal} {Phys. Rev. B}\ }\textbf {\bibinfo {volume}
  {86}},\ \bibinfo {pages} {075128} (\bibinfo {year}
  {2012}{\natexlab{b}})}\BibitemShut {NoStop}%
\bibitem [{\citenamefont {Nolting}\ and\ \citenamefont {Ole\'s}(1980)}]{Nol80}%
  \BibitemOpen
  \bibfield  {author} {\bibinfo {author} {\bibfnamefont {W.}~\bibnamefont
  {Nolting}}\ and\ \bibinfo {author} {\bibfnamefont {A.~M.}\ \bibnamefont
  {Ole\'s}},\ }\href {\doibase 10.1103/PhysRevB.22.6184} {\bibfield  {journal}
  {\bibinfo  {journal} {Phys. Rev.~B}\ }\textbf {\bibinfo {volume} {22}},\
  \bibinfo {pages} {6184} (\bibinfo {year} {1980})}\BibitemShut {NoStop}%
\bibitem [{\citenamefont {Nolting}\ and\ \citenamefont {Ole\'s}(1981)}]{Nol81}%
  \BibitemOpen
  \bibfield  {author} {\bibinfo {author} {\bibfnamefont {W.}~\bibnamefont
  {Nolting}}\ and\ \bibinfo {author} {\bibfnamefont {A.~M.}\ \bibnamefont
  {Ole\'s}},\ }\href {\doibase 10.1103/PhysRevB.23.4122} {\bibfield  {journal}
  {\bibinfo  {journal} {Phys. Rev.~B}\ }\textbf {\bibinfo {volume} {23}},\
  \bibinfo {pages} {4122} (\bibinfo {year} {1981})}\BibitemShut {NoStop}%
\bibitem [{\citenamefont {Bieniasz}\ and\ \citenamefont
  {Ole\'s}(2013)}]{Bie13}%
  \BibitemOpen
  \bibfield  {author} {\bibinfo {author} {\bibfnamefont {K.}~\bibnamefont
  {Bieniasz}}\ and\ \bibinfo {author} {\bibfnamefont {A.~M.}\ \bibnamefont
  {Ole\'s}},\ }\href {\doibase 10.1103/PhysRevB.88.115132} {\bibfield
  {journal} {\bibinfo  {journal} {Phys. Rev. B}\ }\textbf {\bibinfo {volume}
  {88}},\ \bibinfo {pages} {115132} (\bibinfo {year} {2013})}\BibitemShut
  {NoStop}%
\bibitem [{\citenamefont {Berciu}\ and\ \citenamefont
  {Sawatzky}(2009)}]{berciu09}%
  \BibitemOpen
  \bibfield  {author} {\bibinfo {author} {\bibfnamefont {M.}~\bibnamefont
  {Berciu}}\ and\ \bibinfo {author} {\bibfnamefont {G.~A.}\ \bibnamefont
  {Sawatzky}},\ }\href {\doibase 10.1103/PhysRevB.79.195116} {\bibfield
  {journal} {\bibinfo  {journal} {Phys. Rev.~B}\ }\textbf {\bibinfo {volume}
  {79}},\ \bibinfo {pages} {195116} (\bibinfo {year} {2009})}\BibitemShut
  {NoStop}%
\end{thebibliography}%

\end{document}